\documentclass[preprint]{revtex4}

\usepackage{amsfonts,amsmath,amsxtra,graphicx} 

\begin{document}

\title{Subtleties in the quasi-classical calculation of Hawking 
radiation}

\author{Emil T.Akhmedov}
\email{akhmedov@itep.ru}
\affiliation{Moscow, B.Cheremushkinskaya, 25, ITEP, Russia 117218} 

\author{Terry Pilling}
\email{Terry.Pilling@ndsu.edu}
\affiliation{Department of Physics,
North Dakota State University, Fargo, ND 58105} 

\author{Douglas Singleton}
\email{dougs@csufresno.edu}
\affiliation{Physics Department, CSU Fresno, Fresno, CA 93740-8031} 

\date{\today} 

\begin{abstract}
The quasi-classical method of deriving Hawking radiation is investigated. In 
order to
recover the original Hawking temperature one must take into account a 
previously
ignored contribution coming from the temporal part of the action. This 
contribution plus a contribution coming from the spatial part of the action 
gives the correct temperature.
\end{abstract} 

\maketitle

{\bf Introduction:} By considering a quantum field in a fixed
black hole background of mass $M$ one finds \cite{hawking} that
the black hole emits thermal radiation with a temperature $T_H =
\frac{\hbar}{8 \pi M}$ (we set $k_B=c=G=1$). The crucial point
leading to this result is that the collapse of a star into a black
hole changes the vacuum state for quantum fields. The latter
process converts zero modes with respect to the vacuum before the collapse
into a Planckian spectrum with respect to the vacuum after the collapse. This
gives an attractive picture for the origin of black hole
radiation. However, this picture does not give a microscopic
description of the radiation, which would be helpful e.g. for taking
into account the back--reaction of the radiation on the black hole. 

The hope is that a quasi--classical ``tunneling'' process may
give such a microscopic description. The
``tunneling'' picture seeks to describe black hole
radiation by pair creation near the black hole horizon with the
subsequent tunneling of one of the particles of the pair through
the horizon. The tunneling rate is found as the exponent
of the imaginary part of the classical action for the particles
coming from the vicinity of the horizon. 

However, the interpretation of the imaginary
contribution to the particle's action as an indication of
tunneling is subtle. First, if the pair is
created behind the horizon neither of the particles can tunnel
through the horizon, because the tunneling process in quantum
mechanics is described via the solution of a Cauchy problem and has
to be causal, while passing through the horizon is acausal. In 
quantum mechanics the vacuum remains unchanged
which is the reason why we can safely convert a time evolution problem into
an eigen--value problem.
Second, if the pair is created outside a horizon
the time for one of the particles to cross the horizon is
infinite for the stationary distant observer. 
However, this same observer should see the radiation
from the black hole in finite time after the collapse. Finally, the description of
black hole radiation as pair creation in a
strong gravitational field via the production of virtual
super--luminal particles does not have an explicit calculation. 

In \cite{kraus} \cite{parikh} the following idea was proposed to give a physical
picture for the quasi--classical tunneling description of black
hole radiation: during the pair creation process the horizon shrinks slightly
so that the radiated particle already appears outside the horizon. This
obviates the problems discussed in the previous
paragraph. This method of calculating Hawking radiation has also
been shown to give the standard relationship between black hole
temperature and entropy \cite{pilling}. 

In turn the calculation of the imaginary contribution to the
classical action in a black hole background via the solution of
the Hamilton--Jacobi equation is analogous to the quasi--classical
approximation for the calculation of the probability of vacuum
decay in an external electro--magnetic field \cite{schwinger}.
In this type of calculation applied to a gravitational
background one looks for the imaginary contribution to the vacuum
decay amplitude, i.e. Tr$\log{[\Box(g) + m^2]} = \int Dx(t) \,
e^{-\frac{i}{\hbar}\, S(g,x)}$, where $\Box(g)$ is the d'Alembertian
operator in the background metric $g_{\mu \nu}$ for a particle of mass
$m$. On the right hand side we have a path integral over closed paths
and $S$ is the action for particles in the gravitational
field. In the quasi--classical calculation ($\hbar \to 0$) we find
the  saddle--point approximation for the path integral. The main
imaginary contribution comes from the closed paths which cross
the horizon going out and back. In other words in the quasi--classical 
calculation we find an imaginary eigen--value contribution to 
Tr$\log{[\Box(g) + m^2]}$.

{\bf Hamilton-Jabobi Equations:} 
For a particle, of mass, $m$, the Hamilton-Jacobi equation is
\begin{equation}
\label{hj}
g^{\mu \nu} (\partial _\mu S ) (\partial _\nu S) + m ^2 = 0 ~,
\end{equation}
where $g^{\mu \nu}$ is the inverse metric of the background space--time,
$S(x)$ is the action of the particle, in terms of which the
corresponding scalar field can be written as $\phi (x) \propto
\exp [ - \frac{i}{\hbar}\, S(x) + ... ]$. 

For stationary space--times with a time-like Killing vector, to
describe a positive energy state, one can split the action into a
time and spatial part
\begin{equation}
\label{phase}
S (x ^\mu ) = Et + S_0 (\vec{x}),
\end{equation}
where $E$ is the particle energy and $x^\mu = (t, \vec{x})$.
Inserting some particular metric into (\ref{hj}) gives an equation
for $S_0 ( \vec{x})$ which has the solution $S_0 = - \int p_r dr$
where $p_r$ is the canonical momentum in the background metric. If $S_0$ has
an imaginary part one can read off the temperature of the
radiation as follows: The rate for a quasi--classical process can
be written as
\begin{eqnarray}
\label{gamma}
\Gamma &\propto & \exp \left[  - {\rm Im} \oint p_r dr / \hbar \right]
= \exp \left[  - {\rm Im} \left( \int p_r ^{Out} dr - \int p_r ^{In} dr \right) / \hbar \right] \nonumber 
\\
&=& \exp \left[ \mp 2 {\rm Im} \int p_r ^{Out, In} dr / \hbar \right].
\end{eqnarray}
The closed path of integration in the first expression goes across
the barrier {\it and} back (in our case the integration goes
around the horizon in the complex $r$ plane). $S_0 ^{Out, In} =
-\int p_r ^{Out, In} dr$ are the two different traversal directions
across the barrier. The expression in the second line of
\eqref{gamma} assumes that $p_r ^{Out} = - p_r ^{In}$ i.e.
crossing the barrier left to right versus right to left only
differs by a sign. Strictly one should use $\oint p_r dr = \int
p_r ^{Out} dr - \int p_r ^{In} dr$ since only this quantity, {\it
not} $2 \int p_r ^{Out, In} dr$, is invariant under canonical
transformations (see e.g. \cite{chowdhury} for the discussion in
the context of black holes). Thus, only $\oint p_r dr$ is a proper
observable. It happens that in the black hole background for the
metrics which are regular on the horizon Im$\oint p_r dr \neq
\pm 2{\rm Im}\, \int p_r ^{Out, In} dr$, because $\int p_r ^{Out} dr$ has a
non--zero imaginary contribution, while $\int p_r ^{In} dr$ does
not. Below we consider such a difference in greater detail for the
black hole metric in Painlev{\'e} coordinates. 

In order to obtain a temperature from \eqref{gamma} we associate
it with a Boltzmann factor $\Gamma \propto \exp{(-E /T)}$ so that $T
= \frac{ E \hbar}{Im \oint p_r dr}$ (note that $T$ is independent of $E$).

{\bf Spatial Contribution:}
The $1+1$ Schwarzschild metric is
\begin{equation}
ds^2 = - \left(1 - \frac{2M}{r}\right) \, dt^2 +
\frac{dr^2}{\left(1 - \frac{2M}{r} \right)}. \label{schwarz}
\end{equation}
By spherical symmetry the angular part is not important.
The above procedure gives
\begin{equation} 
S_0 = \pm \int_0^{+\infty} \frac{dr}{\left(1 -
\frac{2M}{r}\right)}\, \sqrt{E^2 - m^2 \left(1 
 -\frac{2M}{r}\right)} = \pm \int _0 ^\infty p_r ^{In, Out} dr.\label{intsch}
\end{equation}
It is well known \cite{akhmedov} that doing the contour integration
for (\ref{intsch}) by going around the pole at $r = 2M$ with a semi-circular 
path gives ${\rm Im} (S_0)^{(In, Out)} = \pm 2 \pi M E$ where $+=In$ is for the
ingoing particle and $-=Out$ is for the outgoing particle. This yields a temperature 
twice as large as Hawking's original calculation.
Initially this disagreement between the quasi-classical approach and other 
methods was attributed to the ``badness" of the Schwarzschild
coordinates near the horizon. Below we show that the source of the
disagreement comes from a missed contribution coming from the
temporal part of the total action, $S (x ^\mu) = Et + S_0$. 

Since $\oint p_r dr$ is canonically invariant it will be the same
in any coordinate coordinates related to the Schwarzschild coordinates via a
canonical transformation and one does not need to re-calculate it
in different coordinates. However we mention briefly how
$\oint p_r dr$ remains fixed in two commonly used coordinate
systems: isotropic and Painlev{\'e}-Gulstrand. In isotropic
coordinates, which are related to Schwarzschild coordinates via
$r=\rho(1+M/2\rho)^2$, one apparently finds \cite{vanzo} ${\rm Im} S_0 ^{In, Out} =
\pm 4 \pi M E$ if one treats the isotropic $\rho$ exactly like
$r$.  However near the horizon the relationship between $r$ and $\rho$ goes
like $\rho \simeq \sqrt{r}$ \cite{akhmedov}. One must apply the transformation
to both the integrand {\it and} the contour of the integration in $S_0$. This
transforms the semi-circular contour of the Schwarzschild coordinates to
a quarter circle contour in the isotropic coordinates. Thus one has $i
\frac{\pi}{2} \times Residue$ rather than $i \pi \times Residue$
which again yields ${\rm Im} S_0 ^{In, Out}= \pm 2 \pi M E$ \cite{akhmedov}. In
the Painlev{\'e}-Gulstrand coordinates, which are related to the
Schwarzschild coordinates via  $dt' = dt + \frac{\sqrt{\frac{2M}{r}} \,
dr} {1-\frac{2M}{r}}$ with radial and angular coordinates the
same, one finds that only the outgoing path has an imaginary contribution. 
This happens because a freely falling particle which crosses the horizon does 
not have any barrier in the Painlev{\'e}-Gulstrand coordinates.
This leads to $\oint p_r dr =
\int p_r ^{Out} dr - \int p_r ^{In} dr = \int p_r ^{Out} dr - 0=
4 \pi M E$. Unlike the Schwarzschild or isotropic coordinates,
where both ingoing and outgoing paths contribute equally, here
only the outgoing path contributes with twice the magnitude. The
net result in {\it all} coordinates is ${\rm Im} \oint p_r dr =  4 \pi M E$. Taken by itself this
spatial contribution would give a Hawking temperature twice the
expected value. 

{\bf Temporal Contribution:} The resolution to this factor of two
in the temperature comes from a previously missed, imaginary
contribution from the time part of $S (x ^\mu)$ i.e. from $E t$.
Thus, we really have a quasi--classical process in two coordinates
rather than in one. The reason why there is a non--trivial
contribution to the imaginary action coming from the time part is
because the ``$t$'' coordinate inside the horizon is different
from the ``$t$'' coordinate outside the horizon. To see the
relation between those two ``$t$'' coordinates let us use the
Kruskal-Szekeres coordinates to go across the horizon. 

The Kruskal-Szekeres coordinates $(T,R)$ are related to
Schwarzschild $(t,r)$ in the region exterior to the black hole ($r
> 2M$) given by
\begin{equation}
\label{kruskal1}
T = \left( \frac{r}{2M} -1 \right) ^{1/2} e^{r/4M} \sinh \left( \frac{t}{4M} 
\right) \; , \; \; \;
R = \left( \frac{r}{2M} -1 \right) ^{1/2} e^{r/4M} \cosh \left( \frac{t}{4M} 
\right).
\end{equation}
For the interior of the black hole ($r < 2M$) the relationship is
\begin{equation}
\label{kruskal2}
T = \left( 1 - \frac{r}{2M}\right) ^{1/2} e^{r/4M} \cosh \left( \frac{t}{4M} 
\right) \; , \; \; \;
R = \left( 1 - \frac{r}{2M}\right) ^{1/2} e^{r/4M} \sinh \left( \frac{t}{4M} 
\right).
\end{equation}
To connect these two patches across the horizon at Schwarzschild
$r= 2 M$ one needs to ``rotate" the Schwarzschild $t$ as $t
\rightarrow t - 4 M i \frac{\pi}{2} = t - 2 \pi i M$ (together
with the change $r-2M \to 2M-r$) and then one sees that $(T,R)$
for the exterior patch given in \eqref{kruskal1} becomes the
$(T,R)$ for the interior patch given in \eqref{kruskal2}. This
``rotation'' also needs to be taken into account in the time part
of the action (i.e. in $E t$). This leads to an additional
imaginary contribution coming from the temporal part of the
action. The imaginary
part of this temporal contribution is ${\rm Im} (E \Delta t ^{In, Out}) = - 2 \pi M E$.
For a round trip this will yield ${\rm Im} (E \Delta t ) = - 4 \pi M E$ which is
the same magnitude as the spatial contribution ${\rm Im} \oint p_r dr = 4 \pi M E$.
Thus the total imaginary contribution in any coordinate frame related 
to Schwarzschild by a canonical transformation can be written (remembering that
$S_0 = - \int p_r dr$)
\begin{equation}
{\rm Im} (S (x ^\mu )) = {\rm Im} \left(E \Delta t^{Out}
+ E \Delta t^{In} - \oint p_r dr \right) = -8 \pi M E ~,
\end{equation} 
which yields the canonical Hawking temperature $T_H =  \frac{\hbar}{8 \pi M}$. 

We have shown that the quasi--classical method has a previously
overlooked imaginary part coming from the temporal part of the
total action. Only by taking this extra contribution into account
does one recover the standard Hawking temperature. A similar time
contribution must be taken into account when one applies the
quasi-classical method to other gravitational backgrounds 
(e.g. Reissner-Nordstr{\/o}m, Kerr, Kerr-Newman, deSitter \cite{akhmedova}). Note that in
the case of Unruh radiation there is no contribution from 
the time part since the time coordinate does not have a similar change
when one crosses from one section of space--time to the other.  

{\bf Acknowledgments} ETA acknowledges discussions
with G.Volovik and support from the Agency of Atomic Energy of
Russian Federation and a CSU Fresno Provost Grant.


\begin{thebibliography}{99} 

\bibitem{hawking} S.W. Hawking, Comm. Math. Phys. {\bf 43}, 199 (1975). 

\bibitem{kraus} P. Kraus and F. Wilczek, Nucl. Phys. B {\bf 437} 231 (1995). 
E. Keski-Vakkuri and P. Kraus, Nucl. Phys. B {\bf 491} 249 (1997). 

\bibitem{parikh} M.K. Parikh and F. Wilczek, Phys. Rev. Lett. {\bf 85}, 5042 
(2000); M.K. Parikh Int. J. Mod. Phys. D {\bf 13} 2351 (2004). 

\bibitem{pilling} T. Pilling, Phys. Lett. {\bf B660}, 402 (2008).  

\bibitem{schwinger} J. Schwinger, Phys. Rev. {\bf 82}, 664 (1951). 

\bibitem{akhmedov} E.T. Akhmedov, V. Akhmedova and D. Singleton, Phys. Lett. 
B, {\bf 642} 124 (2006); E.T. Akhmedov, V. Akhmedova, D. Singleton, and T. 
Pilling, Int. J. Mod. Phys. A, {\bf 22} 1705 (2007). 

\bibitem{chowdhury} B. D. Chowdhury, Pramana {\bf 70}, 593 (2008) 

\bibitem{vanzo} M. Angheben, M. Nadalini, L. Vanzo, and S. Zerbini,
JHEP, {\bf 0505:014} (2005) 

\bibitem{akhmedova} 
V. Akhmedova, T. Pilling, A. de Gill, and D. Singleton, 
Phys. Lett. B, {\bf 666} 269 (2008)

\end{thebibliography}
\end{document}